# High pressure bulk synthesis of InN by solid state reaction of binary oxide in a multi-anvil apparatus

*Elena Del Canale,[1,2]\* Lorenzo Fornari,[1,2] Chiara Coppi,[1,2] Giulia Spaggiari,[1,3] Francesco Mezzadri,[2,1] Giovanna Trevisi,[1] Patrizia Ferro,[1] Edmondo Gilioli,[1] Massimo Mazzer,[1] Davide Delmonte.[1]*

[1] CNR – IMEM, 43124, Parma, Italy

[2] SCVSA Department, Università degli Studi di Parma, 43124, Parma, Italy

[3] Department of Mathematical, Physical and Computer Sciences, Università degli Studi di Parma, 43124, Parma, Italy

\*Corresponding Author: elena.delcanale@imem.cnr.it

**Abstract**

We present a new method to synthesize bulk indium nitride by means of a simple solid-state chemical reaction carried out under hydrostatic high pressure/high temperature conditions in a multi-anvil apparatus, not involving gases or solvents during the process. The reaction occurs between the binary oxide $In_2O_3$ and the highly reactive $Li_3N$ as nitrogen source, in powder form. The formation of the hexagonal phase of InN, occurring at 350 °C and P ≥ 3 GPa, was successfully confirmed by powder X-ray diffraction, with the presence of $Li_2O$ as unique byproduct. A simple washing process in weak acidic solution followed by centrifugation, allowed to obtain pure InN polycrystalline powders as precipitate. With an analogous procedure it was possible to obtain pure bulk GaN, from $Ga_2O_3$ and $Li_3N$ at T ≥ 600°C and P ≥ 2.5 GPa. These results point out, particularly for InN, a clean, and innovative way to produce significant quantities of one of the most promising nitrides in the field of electronics and energy technologies.

## 1. Introduction

Indium nitride, as well as its isostructural compound gallium nitride and their solid solutions, is a direct band gap III-V semiconductor of great interest in the field of electronics,[1,2,3] high power devices[4,5,6] and LED technologies,[7,8] for its unique optical absorption properties. From the structural point of view, InN crystallizes in a wurtzite-like hexagonal cell (space group *P6₃mc*) characterized by lattice parameters a = 3.536 Å and c = 5.709 Å.[9] This black, low band-gap semiconductor ($E_g$ = 0.7 eV)[10] shows a notable RT-conductivity, high absorption coefficient, very high thermal conductivity despite an intrinsic chemical instability due to the very weak In-N covalent bond.

InN electro-optical properties completely differ from those of its related compound, GaN, which is instead a white high band-gap semiconductor ($E_g$ = 3.4 eV)[11], looking yellowish due to the presence of defects. Even though isostructural, hexagonal GaN possesses a far smaller unit cell[12] due to the large difference (exceeding 30%) between the two cation ionic radii. As a consequence, GaN is characterized by complementary physical properties i.e., low RT-conductivity, low absorption coefficient and low thermal conductivity.

Due to these opposite characteristics, the stabilization of (In,Ga)N solid solutions, which in principle may allow to access an enormous range electro-optical/thermal properties and get into a wide gamma of technologies, results to be complex, limited at the nanoscale[13,14,15,16,17,18,19,20,21,22] and far from application yet. Consequently, (In,Ga)N production still represents one of the biggest challenges for the scientific community working on this class of nitrides. At the origin of this problem, there is basically the different stability of GaN and InN phases. Indeed, several successful examples of GaN synthesis have been reported in literature, both as film by hetero/homo-epitaxial growth methods (mainly with CVD,[23,24] MOCVD,[25,26,27,28] MBE,[29,30,31,32] MOVPE[33,34,35] and VPE[36]) and as bulk by various crystal growth routes (such as high pressure/high temperature fluid-mediated synthesis methods, e.g. ammonothermal method,[37] high-pressure nitrogen solution growth process[38,39] and Na-flux method under high pressure of the reactive atmosphere[40]) and also via solid state high pressure/high temperature reaction exploiting a cubic anvil cell.[41,42]

Vice versa there are only few reports of effective InN syntheses. This depends on the fact that InN has a low dissociation temperature (450°C [43]), which makes the epitaxial growth at standard temperatures very difficult. Moreover, the synthesis of bulk crystalline InN results even more complex and tricky due to high instability of the In-N bond. In the case of profitable attempts, the reactions often occur under harsh conditions, through the use of toxic, polluting and very hazardous reactants, such as: ammonothermal growth from $InCl_3$ and $KNH_2$ in supercritical ammonia at 2.8 kbar,[44] solvothermal reaction of $InCl_3$/$InI_3$ with $LiNH_2$ in benzene,[45] microwave plasma sources at sub-atmospheric pressure by saturating indium with nitrogen,[46] low-temperature synthesis via nitridation of $LiInO_2$[47] or $In(OH)_3$[48] with $NaNH_2$ flux in autoclave, nitridation of $In_2O_3$ and $In(OH)_3$ with $NH_3$ at 600 °C,[49] solid-state exchange reaction between Ga/$InI_3$ and $Li_3N$[50] or $InBr_3$ and $NaN_3$.[51] Despite the different methods and conditions, the reactions have a very low yield, somehow producing well shaped μm-scale crystals, but rarely a pure bulk product. For these reasons, the use of InN is mainly confined at the lab scale for research purposes, and the known synthesis methods are currently characterized by very high production costs and difficult scalability.

However, some works in literature showed that it is possible to obtain several nitrides also with simpler, cheaper, and scalable techniques, as the case of the mechanochemical (MC) reactions exploited for CrN,[52] $Si_3N_4$,[53] ZrN,[54] GaN,[55] TiN[56] and $Fe_3N_4$.[57] In conclusion, new approaches to synthesize bulk InN are needed to push forward the interest in this class of nitrides and their solid solutions.

Therefore, we started to investigate the MC of InN by applying different ball milling conditions during the treatment. On the basis of this study, the gathered information drove us to identify a new approach for the synthesis of such nitrides, by the use of hydrostatic high pressure/high temperature (HP/HT) reactions performed in a multi-anvil press. To the best of our knowledge, some nitrides have already been studied by analogous HP/HT methods, as in the case of AlN,[58] $SrN_2$ and $\gamma$-$P_3N_5$,[59] $Ge_3N_4$[60] by multi-anvil, c-$BC_2N$,[61] $B_{13}N_2$[62] by toroid-type high pressure apparatus, MoN and $MoN_{1-x}$,[63] $c$-$Si_3N_4$,[64] PtN[65] by diamond anvil cell and $NaN_3$[66] by Merrill-Bassett type pressure cell. However, the only attempts to synthesize InN in HP/HT conditions (i.e., in the giga-Pascal pressure range, exceeding the kbar regime typical of the solvothermal methods previously reported), again led to the formation of tiny crystals from the direct synthesis of metallic indium ad compressed nitrogen at 2 GPa and 700°C,[67] but neither exploiting simple solid-state reaction nor applying pressure with solid media.

In this paper, we show that a solid-state reaction under isotropic HP/HT conditions leads to the formation of pure bulk InN, starting from indium binary oxide and non-toxic nitrogen-based compounds, without the use any organic solvents or gases during the process. The relatively mild conditions enable to obtain a significant amount of material, with high yield. The same procedure was then successfully applied for the synthesis of pure bulk GaN, finding the stability condition for temperatures higher than InN.

## 2. Experimental Methods

The mechano-chemical reactions (MC) were performed using a Pulverisette 7 Classic Line high energy planetary ball mill (Fristch GmbH), with sealed $ZrO_2$ jars (volume: 45 ml) and spheres (diameter: 10 mm). $In_2O_3$ (ChemPur, 99.99%) and a super-stoichiometric content (i.e., +50% of excess) of $Li_3N$ (Alfa Aesar, 99.4%) were used as precursors and mixed under inert atmosphere. The reactions were carried out under a controlled inert atmosphere in dry conditions, thus without the use of an assisting liquid, inserting the ball mill apparatus in a glove box with slight $N_2$ overpressure, varying the combination of the main MC parameters: rotational speed (RPM), ball-to-powder mass ratio (BPR) and reaction time (TIME).

High pressure high temperature (HP/HT) syntheses were performed in isotropic conditions through a multi-anvil 6/8 Walker-type press apparatus (Rockland Inco Corps.). The reaction was carried out starting from a 600-700 mg homogeneous mixture of powders of $In_2O_3$ (ChemPur, 99.99%) with a super-stoichiometric content (i.e., +50% of excess) of $Li_3N$ (Alfa Aesar, 99.4%) in an Au capsule . Once the target pressure (in the 2.5 – 6 GPa range) was reached with a ramp rate of about 30kPa/min, the temperature was increased of 50°C/min until the desired value (in the 350 – 900 °C range). After the synthesis, the temperature was quenched down to room temperature and the pressure slowly released overnight. The products, obtained in form of dense cylinders (about 5x5x5 mm), are ground in an agate mortar or cut in form of discs to be further characterized.

The phase analysis was performed through Powder X-Ray Diffraction (PXRD), using two different diffractometers in Bragg-Brentano geometry: (I) for qualitative analysis of the products, a Thermo-Electron X'Tra diffractometer equipped with a Thermo Electron solid state Si(Li) detector was used. This instrument utilizes Cu-K$\alpha$ wavelength ($\lambda$=1.5406 Å) and data were collected with 0.05° step and 3s of counting time; (II) For quantitative data collection we exploited a Rigaku Smartlab XE diffractometer with Cu $K_\alpha$ wavelength. A Ni filter was used to suppress the $K_\beta$ contribution. 5.0° soller slits were located both on the incident and diffracted beam and data were collected using a HyPix3000 detector operating in 1D mode.

The morphology and composition of the samples were investigated with a Zeiss Auriga Compact Field-Emission Scanning Electron Microscope (SEM) equipped with an Oxford Xplore 30 Energy Dispersive Spectroscopy (EDS) system. SEM images were acquired by using both 20 kV and 5 kV acceleration voltages of the primary electron beam. EDS analyses were conducted by exciting the samples with a 20 kV accelerated electron beam.

Raman measurements were carried out using a micro-Raman spectrometer (Horiba LabRam HR Evolution Raman) equipped with a confocal Olympus microscope and 10x, 50x, ULWD50x, 100x objectives (spatial resolutions of approximately 1 μm). The Micro-Raman apparatus is completed by a He-Ne laser emitting at 632.8 nm, BraggRate Notch Filters, Silicon CCD + InGaAs Diode Array detectors, gratings 300-600-1800 lines/mm, and density filters. The spectrometer was calibrated using the standard silicon Raman peak at 520.6 cm-1 before each measurement. The spectra here reported were recorded using the 100x objective, for 30 s and 4 repetitions.

## 3. Results and Discussion

The solid-state chemical reaction is a nitridation synthesis of the binary In oxide by the use of $Li_3N$ (in super-stoichiometric ratio, 50% excess in order to balance the eventual weighting error coming from conspicuous water absorbance during the alkali nitride precursor preparation), as reported in equation (1):

$$In_2O_3 + 2\ Li_3N \rightarrow 2\ InN + 3\ Li_2O \tag{1}$$

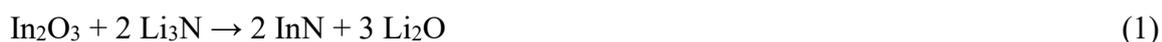

### 3.1. Mechanochemistry of InN

A mechanochemical process exploits the non-equilibrium thermodynamics and very high local temperatures differently from conventional synthesis methods, accessible via high energy ball milling experiments. The application of this technique led, in the current case, to a better understanding of the InN formation process and the energy needed to initiate and complete the solid state nitridation reactions of equation (1), even though the MC regimes explored in this work did not allow to obtain InN.

In this study, we investigate a wide range of MC parameters exploring two complementary energy regimes, as comprehensively described in paragraph **S.1** in the Supporting Information. It is found that, for low MC energies (400 < RPM ≤ 600), and independently from the ball-to-powder ratio and milling time, it is not possible to activate the nitridation reaction; the process indeed leads to a reduction of the mean crystallinity of the $In_2O_3$ phase (see **Figure S2** in the Supporting Information). Surprisingly, for high energies (RPM > 600), the lithium ions start to react with the oxygens of $In_2O_3$; however, the local mechanical energies are too high to metastabilize In-N bonds and consequently an unwanted redox process become favored: nitrogen ions oxidize, forming molecular gas by forcing $In^{3+}$ reduction to metallic indium (**Figure 1** here below and **Figure S1** in the Supporting Information).

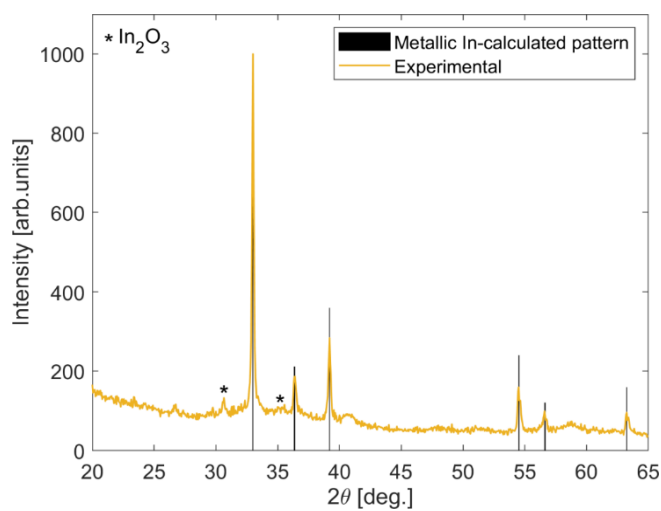

**Figure 1.** PXRD of the MC products obtained for the InN reaction by using 630RPM, 268 BPR and 40min of milling (yellow curve). The black vertical lines represent the calculated pattern of metallic In from ICSD using POWD-12++ 539, 3(1997), while the black "*" symbol represents the calculated $In_2O_3$ pattern from ICSD using POWD-12++1928, 1 (1997).

In brief, high energy ball milling processes seem to hinder the formation of InN so that the wanted reaction is quite inaccessible by MC, at least for an experiment conducted in $N_2$ atmosphere and dry conditions. These results suggest the use of a complementary approach for the study of this nitridation reaction. Therefore, a novel approach, consisting in the combination of high temperatures and high pressures through equilibrium thermodynamic solid-state reaction, is exploited to metastabilize the very weak In-N bond.

### 3.2. HP/HT InN syntheses

The same solid-state nitridation reaction was performed in the HP/HT regime, exploring different conditions of hydrostatic pressure (3 to 6 GPa), temperature (350 to 900°C) and duration (2 to 6 h), as summarized in **Table 1**. It was observed that the HP/HT synthesis of InN requires temperatures below 600 °C to prevent the formation of metallic In (**Figure S3**, blue curve in the Supporting Information) in close analogy to what observed with MC in high RPM regime. Particularly, the polycrystalline hexagonal InN was detected to form around 350 °C (yellow and purple curve **Figure S3**, in the Supporting Information).

PXRD analysis (**Figure S3** in the Supporting Information), highlights the systematical presence of ternary Li-based compounds, coming from a contamination of the highly hygroscopic $Li_3N$ reactants (**Figure S8** in the Supporting Information), that can be almost completely removed with a washing treatment in acidic water but with a very low final mass yield of the InN product (**Figure S4** in the Supporting Information).

**Table. 1**. InN explored HP/HT synthesis conditions.

| Sample | Pressure [GPa] | Temperature [°C] | Time [h] | Nitride presence | Spurious In phases |
|---|---|---|---|---|---|
| InN | 6 | 900 | 2 | NO | YES |
| InN | 3 | 750 | 2 | NO | YES |
| InN | 3.5 | 600 | 3 | NO | YES |
| InN | 6 | 390 | 6 | YES | YES |
| InN | 3 | 350 | 6 | YES | YES |
| InN* | 3.5* | 350* | 6* | YES | NO |

*\* Reaction performed starting with new and pristine $Li_3N$ reagents*

The complete double exchange reaction (see equation (1)) is obtained through a HP/HT synthesis at 3.5 GPa, 350 °C and 1 h (**Figure 2**, upper panel), starting from a pristine $Li_3N$ bottle. Noteworthy, besides InN, only $Li_2O$ is present, as expected.

The sample powders are poured in acidic water (0.1M HCl) and then centrifuged at 4000 RPM for 15 min to separate InN from the other components, which are fully soluble in water. When $Li_2O$ is put in water the solution pH rises, due to the formation of LiOH. To successfully remove all the amount of $Li_2O$ the non-soluble suspension must be washed, centrifuged, and then separated from the aqueous solution several times, until neutral pH is reached. The precipitate is finally dried on a heating plate at 130 °C.

The PXRD results, reported in **Figure 2** (bottom panel), show the complete removal of the secondary phase, leaving a pure InN polycrystalline powder.

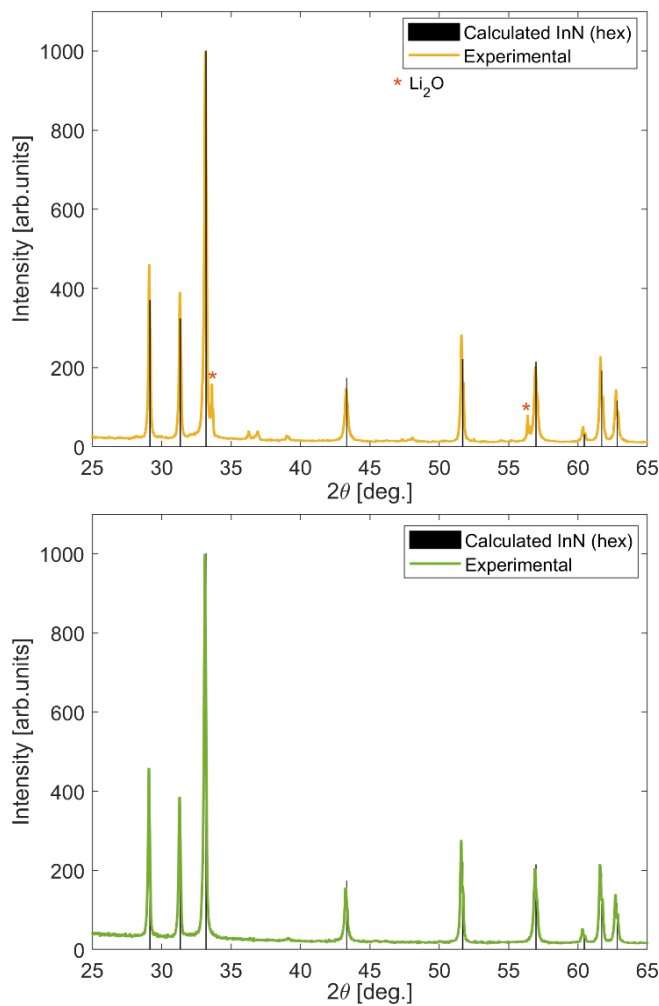

**Figure 2.** PXRD pattern collected for InN after the HP/HT synthesis (upper panel) and after the subsequent washing treatment (bottom panel), obtained at 3.5 GPa and 350°C. The black lines are the calculated reflections of the InN hexagonal phase calculated pattern from ICSD using POWD-12++ 46, 10086 (1997), while the orange "*" symbol correspond to the Li$_2$O phase, calculated from ICSD using POWD-12++40, 588 (1997).

SEM/EDX measurements (**Figure 3**) show that InN crystallites have sub-micrometrical dimensions around 100-200 nm, probably related to the reduced synthesis temperature, not allowing an effective sintering process. The effectiveness of the washing treatment in removing Li-based compounds from the HP/HT synthesized InN is confirmed by the compositional analysis: In and N are estimated with the expected ratio of 1:1, within the instrumental error. The C signal possibly comes from the carbon tape on which the powders are dispersed on the SEM probe, and the minor oxygen contribution could be ascribable to organic contamination on the surface.

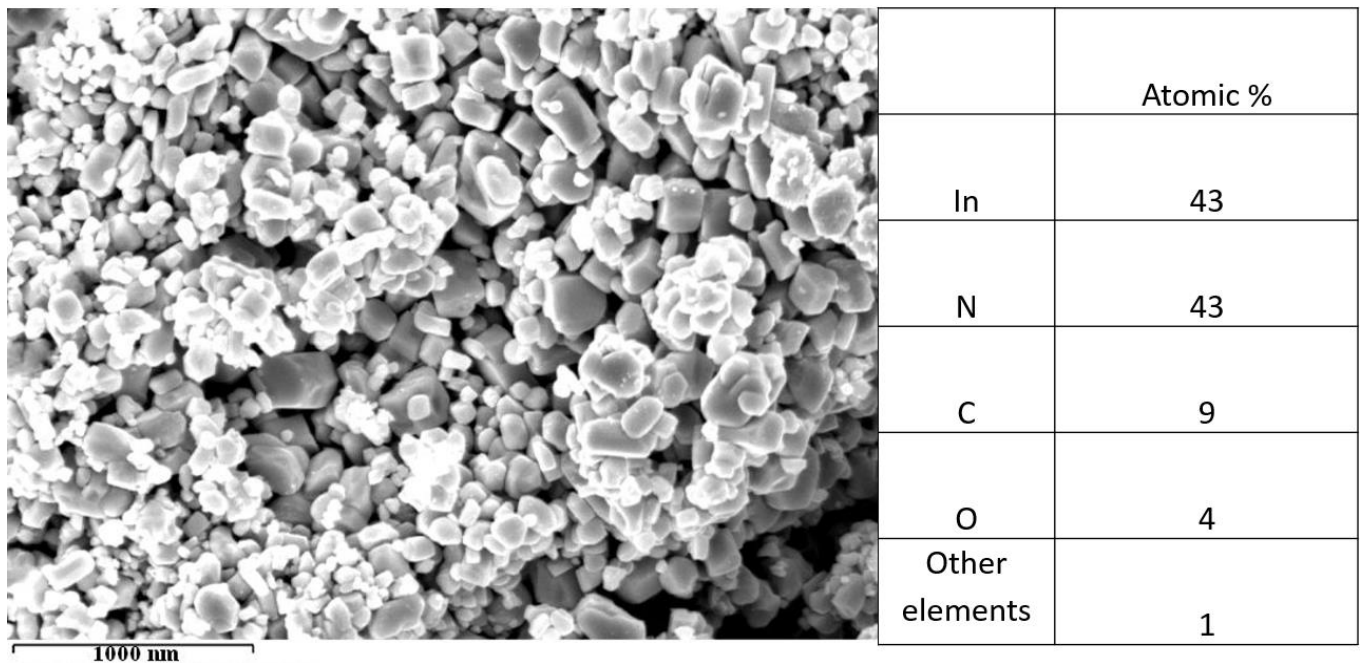

|  | Atomic % |
|---|---|
| In | 43 |
| N | 43 |
| C | 9 |
| O | 4 |
| Other elements | 1 |

**Figure 3.** SEM image with 50kX magnification collected on hexagonal InN crystallites obtained at 350°C, 3 GPa, 1 h and relative atomic compositions.

The Raman spectra of InN (**Figure 4**) presents a noisy signal, ascribable to the observed relatively low crystallinity. However, the main peaks of the hexagonal phase[68] are still identifiable: 480 cm$^{-1}$ A$_1$ (TO), 476 cm$^{-1}$ E$_1$ (TO) and 580 cm$^{-1}$ A$_1$ (LO).

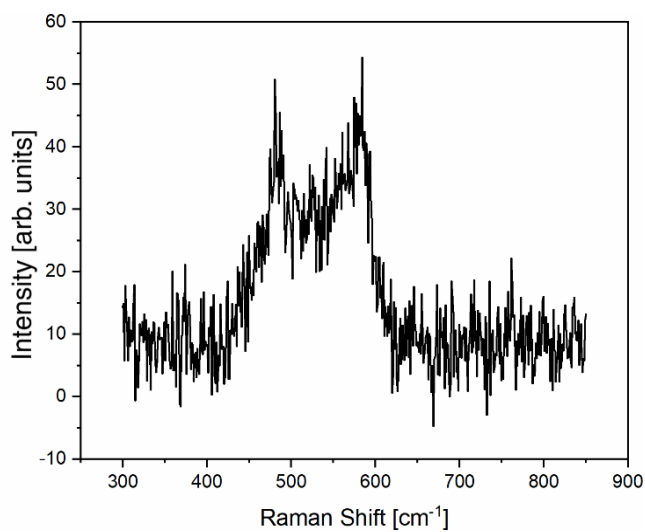

**Figure 4.** Raman spectra of InN powders after the washing treatment.

### 3.2.1 HP/HT synthesis of GaN

Analogous double-exchange reaction has been then studied for the synthesis of bulk GaN to find possible overlapping thermodynamic regimes in which both GaN and InN can be stabilized.

The solid-state chemical reaction proposed for the GaN synthesis is the nitridation of the binary Ga oxide by again the use of Li$_3$N (in super-stoichiometric ratio, 50% excess as for InN), as reported in equation (2):

$$Ga_2O_3 + 2\ Li_3N \rightarrow 2\ GaN + 3\ Li_2O \qquad (2)$$

The results here obtained are resumed in **Table S1** while PXRD data are plotted in **Figure S5** in the Supporting Information. it was observed that the GaN hexagonal phase forms at T ≥ 600 °C and P ≥ 2.5 GPa. As for InN HP/HT synthesis, spurious Li-based binary and ternary oxides are always present and a similar washing treatment was applied to successfully remove the spurious phases (**Figure S6** in the Supporting Information).

The complete double exchange reaction (see equation (2)) can be similarly obtained exploiting pristine Li$_3$N powders, after a HP/HT synthesis performed at 3.5 GPa, 900 °C and 3 h (**Figure 5**, upper panel). The PXRD pattern shows the presence of a pure polycrystalline GaN bulk product (**Figure 5**, bottom panel), except for few peaks with almost negligible intensity (around 33, 38 and 52°) belonging to one or more unidentified phases whose assignation is prohibitive. Raman spectroscopy data and SEM-EDX analysis confirm the obtainment of hexagonal GaN powder with the right composition of mean grain of the crystallites of about 1 μm as shown in **Figure S7** (right and left panel, respectively) in the Supporting Information.

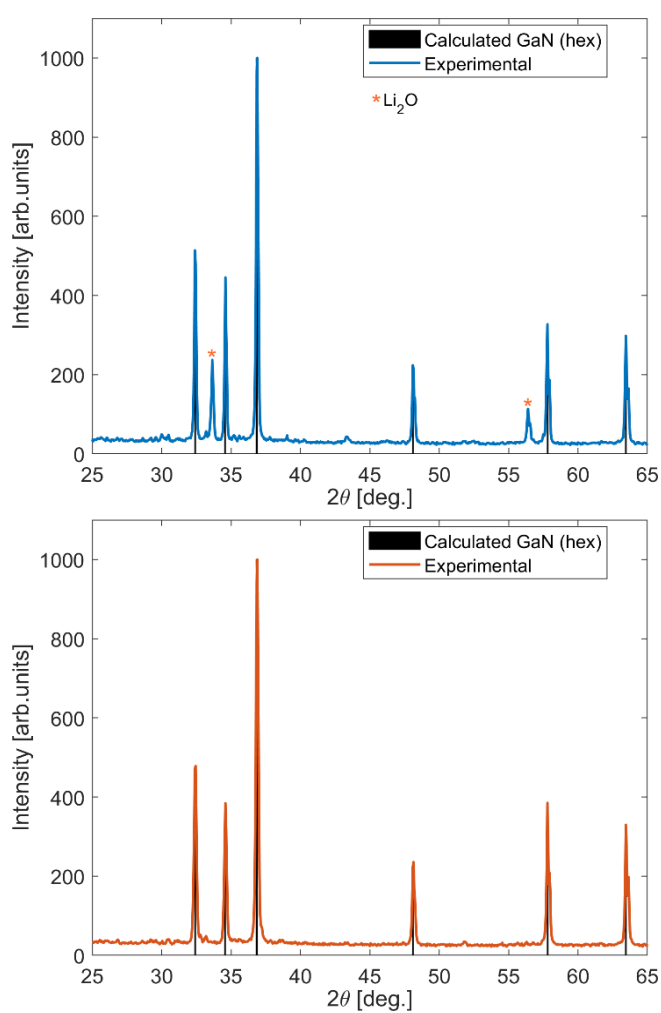

**Figure 5.** PXRD pattern collected for GaN after the HP/HT synthesis (upper panel) and after the subsequent washing treatment (bottom panel), obtained at 3.5 GPa and 900°C. The black lines are the calculated reflections of the GaN hexagonal phase from ICSD using POWD-12++ 23, 815(1997), while the orange "*" symbol indicate to the expected reflections of $Li_2O$, calculated from ICSD using POWD-12++40, 588 (1997).

Noteworthy, from these data it is clear that HP/HT technique has demonstrated to be an effective way to obtain pure bulk GaN in the same pressure regime exploited for the InN synthesis (i.e. P about 3 GPa), even though the temperature conditions required for the stabilization of the two nitrides seem to be incompatible, at least exploiting this particular solid-state reaction and these thermodynamic parameters.

## 4. Conclusions

In this work we reported a comprehensive study of the solid state nitridation reaction of InN from the binary oxide and $Li_3N$ as nitrogen source, without the use of any toxic solvents or gases, following two unconventional synthesis approaches: mechanochemistry (MC) and High Pressure/ High Temperature (HP/HT) synthesis.

MC reaction via high energy planetary ball milling in dry conditions pointed out that InN cannot be obtained because, at the activation threshold of the $Li_3N$-$In_2O_3$ reaction, the In-N bond is not stable, leading to an unwanted redox process with the formation of reduced metallic In and $N_2$. This finding suggested the application of high pressure as a viable way to guide the metastabilization of InN bond.HP/HT synthesis performed in a multi-anvil apparatus effectively allowed toperform the double exchange nitridation reaction, forming hexagonal InN polycrystalline powders and $Li_2O$ as unique byproduct. The reaction succeeded in a wide pressure range (between 3 and 6 GPa) between 350 and 400°C. A simple washing treatment in acidic water allows to separate the InN polycrystalline powder from the soluble $Li_2O$. The same method, applied for the synthesis of GaN, ,enables the analogous double exchange-reaction in the same HP regime but at higher temperatures.

## 5. Supporting Information

The Supporting Information is available free of charge at *link?*

Details on mechanochemistry of InN, additional PXRD measurements on InN synthetized by HP/HT, recap table on the GaN HP/HT syntheses and their structural characterization, SEM and Raman analysis on GaN HP/HT products, PXRD pattern of $Li_3N$ contaminated reactant" (DOC).

## 6. Acknowledgment

This work has been carried out in the framework of the research project BEST4U-Technology for high Efficiency 4 terminal Bifacial Solar Cells for Utility scale (n. ARS01_00519), funded by Programma PON <<R&I>> 2014-2020 –Azione II and it has benefited from the equipment and framework of the COMP-HUB Initiative, funded by the 'Departments of Excellence' program of the Italian Ministry for Education, University and Research (MIUR, 2018-2022) and of the Bio-MoNTANS project, funded by Fondazione Cariparma. L.F., C.C., E.G. and D.D. sincerely thank Prof. Stefano Poli, (Department of Earth Science, University of Milan), for his technical help and advice on the high-pressure synthesis and calibration.

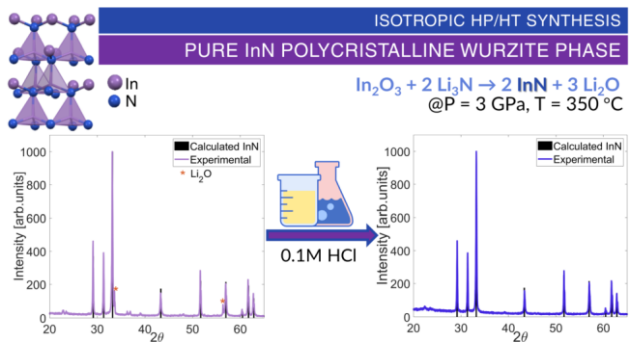

For Table of Contents Only

## Synopsis

Bulk indium nitride was successfully grown by means of a simple solid-state chemical reaction carried out under hydrostatic high pressure/high temperature conditions in a multi-anvil apparatus, starting from the binary oxide and Li$_3$N as nitridation agent, not involving gases or solvents during the process. Simple washing treatment in acidic water allows to remove the unique byproduct (Li$_2$O) from the sample, obtaining pure polycrystalline indium nitride hexagonal phase.